\documentclass{aa}  
\usepackage{graphicx}
\usepackage{txfonts}
\begin{document}
   \title{Particle trajectories and acceleration during 3D fan reconnection}
   \titlerunning{Trajectories during 3D fan reconnection}

   \author{S.~Dalla\inst{1,2} \and P.K.~Browning\inst{1}}

   \authorrunning{Dalla \& Browning}


   \institute{School of Physics and Astronomy, University of Manchester,
              Manchester M13 9PL, UK
            \and
              Centre for Astrophysics, University of Central Lancashire,
              Preston PR1 2HE, UK;
              \email{sdalla@uclan.ac.uk}
             }

   \date{Received}

 
  \abstract
   {The primary energy release in solar flares is almost certainly due to magnetic 
reconnection, making this a strong candidate as a mechanism for particle
acceleration. While particle acceleration in 2D geometries has been widely
studied, investigations in 3D are a recent development. Two main classes of 
reconnection regimes at a 3D magnetic null point have been identified: 
fan and spine reconnection}
   {Here we investigate particle trajectories and acceleration during
  reconnection at a 3D null point, using a test
particle numerical code, and compare the efficiency of the fan and spine
regimes in generating an energetic particle population.}
   {We calculated the time evolution of the energy spectra. We discuss the geometry 
of particle escape from the two configurations and characterise the trapped and
escaped populations.}
   {We find that fan reconnection is less efficent than spine reconnection
in providing seed particles to the region of strong electric field where
acceleration is possible.
The establishment of a steady-state spectrum requires
approximately  double the time in fan reconnection.  
The steady-state energy spectrum at
intermediate energies (protons 1 keV to 0.1 MeV) is comparable in the fan and 
spine regimes. 
While in spine reconnection particle escape takes place in two 
symmetric jets along the spine, in fan reconnection no jets are produced and
particles escape in the fan plane, in a ribbon-like structure.
 }
   {}


   \maketitle
%

\section{Introduction}

%

In recent years, magnetic reconnection has been recognised as the key
process underlying energy release in solar flares.
At the same time, the importance of generalising
the well established 2D reconnection theory to 3D configurations has been
emphasised by many authors. Reconnection in 3D can take place in various locations,
including separator lines; but in initial studies, it is natural to focus on the simplest configuration,
which is also the closest analogue of the classic 2D X-point, namely a 3D magnetic null.
Such null points are likely to be common in the solar corona, due to the complex
topology arising from the mixed polarity photospheric flux sources [\cite{Sch2002,Lon2003}].
The SOHO EIT images have given direct evidence of a 3D null in a solar coronal Active Region [\cite{Fil1999}] and
reconstructions of the solar magnetic fields
of active regions have shown that reconnecting 3D nulls can be present
during solar flares [\cite{Aul2000,Fle2001}]. Recently, Cluster has
found a 3D null point in a reconnection event in the Earth's magnetotail [\cite{Xia2006}].

One important consequence of reconnection is the acceleration of charged particles
by the strong electric fields associated with the process. As confirmed by recent results from RHESSI
[\cite{Lin2003}], 
large numbers of  nonthermal electrons and protons are observed in solar
flares.
It is an attractive hypothesis that magnetic reconnection, the 
primary source of the energy release, is also responsible for generating these high energy 
particles. Particle acceleration by reconnection has been extensively studied in 2D, and 2D
models are continously being improved
(see \cite{Han2006} and references therein). However, very little is known about particle acceleration in 
3D geometries, although these should be common in nature, and the idealisations of 2D models may 
lead to unrealistic results (for example, a 2D configuration has an electric
field of infinite spatial extent in the invariant direction along a line of points
in which the magnetic field is zero). 
Some interesting preliminary studies have been undertaken of test particles
using  3D MHD codes to provide background 
electromagnetic fields [\cite{Sch1999,Tur2005, Tur2006}],  but these are limited by numerical resolution
and it is difficult to derive an understanding of scaling or general properties. 
Recently, we began to study the fundamental properties of particle acceleration
in 3D by considering a basic 3D null point geometry [\cite{Dal2005,Dal2006}]. 

Two regimes of reconnection at a 3D magnetic null are possible: spine and fan
reconnection [\cite{Lau1990,Pri1996}]. While the magnetic field
configuration is the same in the two regimes, they are characterised by
different plasma flow patterns and different electric fields. Spine reconnection has
a current concentration along the critical spine field line, while fan reconnection has a 
current sheet in the fan plane. 

In previous work we 
focussed on the spine reconnection regime, and analysed the
trajectories of single particles, as well as the evolution of 
the spectrum and spatial distributions of a population of 
particles [\cite{Dal2005,Dal2006}]. We found that the 3D spine 
reconnection regime can be an efficient particle accelerator    
and that particles escaping from the region near the null, do
so in two symmetric jets along the spine.

In this paper we focus on the 3D fan reconnection regime, with
specific interest in comparing the efficiency of fan and 
spine reconnection in providing seed particles for acceleration
to the regions of strong electric fields, and in studying the 
3D geometry of particle escape after acceleration. 
We study this problem within the test particle approach, by means of a
numerical code in which the trajectories of a particle population are
numerically integrated. 

We adopt the \cite{Pri1996} model of fan reconnection. 
This model applies to the outer ideal reconnection region 
and does not include acceleration due to the parallel electric field in the
inner dissipation (resistive) region; indeed, the mathematical model breaks down
very close to the spine/fan as the electric field formally diverges (in our 
code, we eliminate this singularity as described in Sec.~2 below).
Despite its limitations, there are advantages to having a relatively simple
field model which highlights key topological and geometrical properties;
the Priest and Titov (1996) model is the closest 3D analogue to the constant 
out-of-plane electric field ($E_z$) which is
very widely used in 2D particle acceleration models [eg: \cite{Dee1991,Vek1997,
Fle1997,Han2006}].
The use of a constant $E_z$ also
excludes the direct parallel electric field  acceleration  (at least in the
case of zero guide field) and is not a self-consistent
reconnection solution in general.
It is important to understand the 3D analogue of the well studied 2D problem,
before investigating more complete but inevitably much more complex models.
Moreover 2D models with more complex fields show good agreement with simplified
models based on inflow to neutral point.

It should also be noted that in a highly conducting plasma, 
the size of the dissipative region will be very small in spatial extent.
The majority of particles in a volume that includes a dissipation region
(a current sheet at the fan plane in fan reconnection) 
will not enter such a region at all. 
We plan to include dissipative effects in future work. Our work here
is thus complementary to that of \cite{Lit2006}, who considers, using an approximate analytical
approach, only the acceleration of particles due to the parallel electric field within the fan plane current
sheet. As has been shown by \cite{Vek1997}, electric fields outside the
dissipation region contribute to the acceleration as well as determining
the way in which particles are supplied to the small resistive region.
The full trajectory approach that we use (as opposed to the guiding center approach) can 
properly describe the acceleration that occurs as particles reach the
region in the vicinity of the fan plane region where the electric field is strong and the 
guiding center approximation breaks down. 

The reconnection configuration is outlined in Section 2. 
Results are presented in Sections 3 and 4, which consider respectively single 
particle trajectories and distributions of particles. The conclusions
are presented in Section 5.


\section{Fan reconnection configuration}\label{sec.config}

We consider a magnetic configuration of the form:
\begin{eqnarray} \label{eq.magnfield}
B_x & = &  B_0 \, x \label{eq.bfan1}\\  
B_y & = &  B_0 \, y \label{eq.bfan2}\\
B_z & = &  - 2 \,  B_0 \, z \label{eq.bfan3} 
\end{eqnarray}
describing a
potential 3D magnetic null, where the spatial variables $x$, $y$ and $z$
are normalised with respect to a length $L$, the size of the simulation
  box, which represents a global length scale. Here $B_0$ is the magnitude of ${\bf B}$ at
$\sqrt{(x^2+y^2)}$=1 and $z$=0. It should be noted that this current-free
null is the simplest of a family of 3D nulls (\cite{Par1996}), but it is used because 
analytical expressions for the electric field in a reconnection situation
are available, as discussed below.  For our potential magnetic
null, an axis of symmetry of the magnetic field, called
the spine, is present, labelled here as the $z$ axis. This critical 
fieldline connects to the null point.  In the plane $z$=0,
the magnetic field lines are straight lines through the null point,
describing a fan. Hence this plane is called the fan plane. The spine and the fan are 
the 3D analogues of the separatrix planes in  classic 2D X-point geometry.

   \begin{figure*}
   \centering
   \includegraphics[width=14cm]{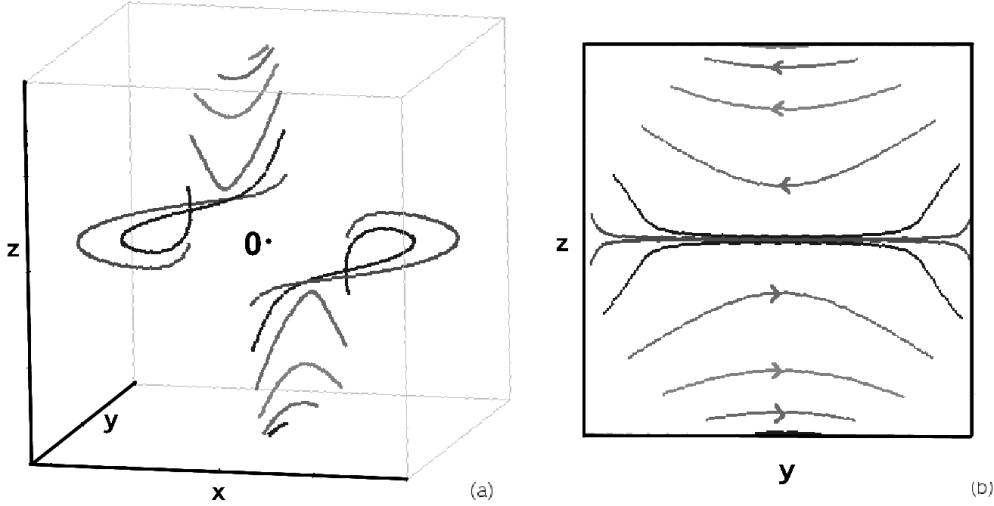}
   \caption{Diagrams of flow lines for 3D fan reconnection: (a) 3D graph of flow lines;
     (b) Projection of the flow lines onto the $y$-$z$ plane.}
              \label{fig.fanflow}
    \end{figure*}

Reconnection regimes at a 3D magnetic null have been described by
\cite{Lau1990} and \cite{Pri1996}.
Two types of reconnection are possible: spine and fan reconnection. We have previously studied
the spine mode [\cite{Dal2005,Dal2006}], hence the focus here is on fan reconnection. For a potential null, the spine
reconnection regime is characterised by plasma flows lying in planes
containing the spine, while fan reconnection has non-planar flows that
carry the magnetic field lines in a \lq swirling\rq-like motion
anti-symmetric about the fan plane. 
The fan plane is thus a singular surface as the flow is discontinous. 
The specific flow pattern of the model we use, results from having imposed, on
surfaces at $z$=$+$1 and $z$=$-$1, that the plasma flow is along straight lines
parallel to the to the $y$-axis and its magnitude only depends on $y$ [\cite{Pri1996}]. In
other words, on the chosen
surfaces: $v_x$=$v_z$=0 and $v_y$=$v_y(y)$, with $v$ the plasma flow velocity.
Note that the resulting electric field and flow pattern are
non-axisymmetric, and that these are not unique, but only one of the 
possible solutions that give rise to fan reconnection.
Imposing the above flows introduces a preferential direction, the $y$ axis
in our geometry, and this will control the particles' behaviour, as shown in
Sections \ref{sec.singlepart} and  \ref{sec.manypart}. Other types
of fan reconnection will similarly be characterised by a preferential
direction, so in this sense our results are generic.

The electric field we consider was derived by \cite{Pri1996} and
has the expression:
\begin{eqnarray}
E_x & = & E_0 \: \frac{- \,|z|^{1/2}}{(4+y^2\,|z|)^{1/2}} \label{eq.efan1}\\
E_y & = & E_0 \: \frac{xy \,|z|^{3/2}}{(4+y^2\,|z|)^{3/2}}  \label{eq.efan2}\\
E_z & = & E_0 \: \frac{-2x \: \mbox{sign}\,z \:|z|^{-1/2}}{(4+y^2\,|z|)^{3/2}}  \label{eq.efan3}
\end{eqnarray}
where $\mbox{sign}\,z $=1 if $z$$>$$0$ and
$\mbox{sign}\,z $=$-$1 if $z$$<$$0$.

Figure \ref{fig.fanflow} shows diagrams representing the plasma flow lines
during  fan reconnection in the Priest and Titov model. Figure \ref{fig.fanflow}-(a) gives a 3D view
of flows for plasma parcels starting in a plane containing the spine, at a
value of the azimuthal angle $\phi$=100$^{\circ}$; Figure
\ref{fig.fanflow}-(b) shows a projection of the flow lines onto the 
$y$-$z$ plane. The azimuthal angle (longitude) is defined as
  $\phi$=$\tan^{-1}(y/x)$, and the latitude  $\beta$ as $\beta$=$\tan^{-1}(z/\sqrt{x^2+y^2})$.

The electric field given by Eqs.~(\ref{eq.efan1})--(\ref{eq.efan3}) is
characterised by a singularity in the fan plane, i.e.~ when $z$=0. We
eliminate this singularity by replacing, in Eq.~(\ref{eq.efan3}):
$ |z|^{-1/2}$  by $ z \,(z^2+l^2)^{-3/4}$,
with $l$ a small constant giving
the length scale of the dissipative layer in which resistivity (or other dissipative effects)
are significant. Hence $l$ can be taken to represent the width of the current
sheet.
In future, the singularity will be
resolved more realistically by using self-consistent reconnection models
incorporating a localised dissipation region at the fan plane.

\section{Single particle trajectories}\label{sec.singlepart}

We first analyse trajectories of a single particle in the 3D fan reconnection 
configuration, by numerically integrating the equations of motion with the
electric and magnetic fields given by
eqs.(\ref{eq.bfan1})--(\ref{eq.efan3}). We solve the relativistic equations of motions
given by:
\begin{eqnarray}  
\frac{d {\bf x}}{d t}  &=& \frac{{\bf p}}{m_0 \gamma} \\
\frac{d {\bf p}}{d t}  &=& q \, \left( {\bf E} + \frac{1}{c}  \frac{{\bf p}}{m_0 \gamma} \times {\bf B} \right)
\end{eqnarray}  
where $t$ is time,
${\bf x}$ and ${\bf p}$ are the particle's position and momentum, $q$ and
$m_0$ its charge and rest mass, $\gamma$ its relativistic factor, and
$c$ is the speed of light.

   \begin{figure*}
   \hspace*{-1.5cm}
   \centering
   \includegraphics[width=14cm]{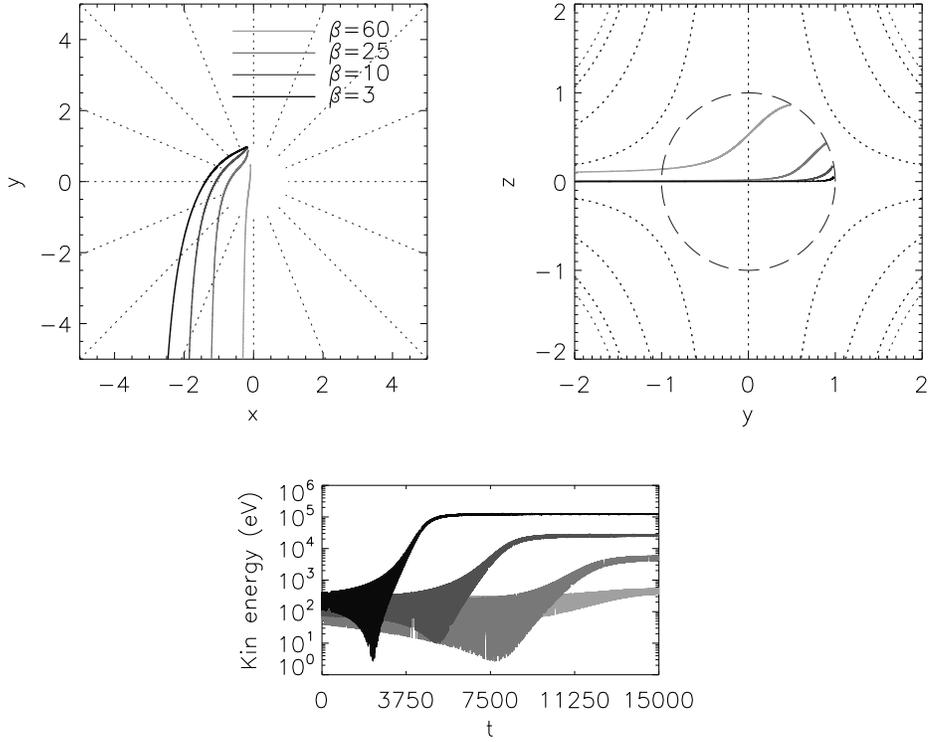}
   \caption{Single particle trajectories in the 3D fan reconnection
     configuration, for initial positions at $r$=1, longitude $\phi$=100$^{\circ}$ and
     latitudes $\beta$=60$^{\circ}$, 25$^{\circ}$, 10$^{\circ}$, 3$^{\circ}$. The initial particle energy is 300 eV. 
     The particle is a
    proton with initial pitch angle 92$^{\circ}$. We use $E_0$=3
     kV/m, $B_0$=100 gauss, L=10 km, t$_{final}$=15000.
Dotted lines indicate projections
of magnetic field lines.}
              \label{fig.varybeta}
    \end{figure*}

   \begin{figure*}
   \hspace*{-1.5cm}
   \centering
   \includegraphics[width=14cm]{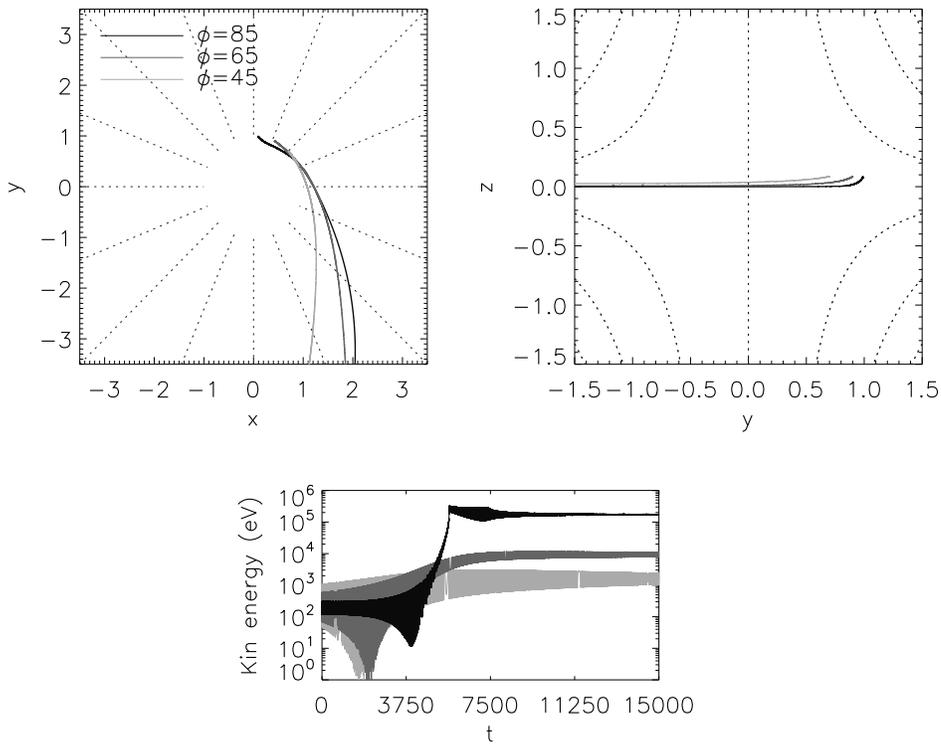}
   \caption{Single particle trajectories in the 3D fan reconnection
configuration, for initial positions at $r$=1, constant latitude $\beta$=5$^{\circ}$
and azimuthal angles  $\phi$=85$^{\circ}$, 65$^{\circ}$, 45$^{\circ}$. The
other input parameters are the same as in Figure \ref{fig.varybeta}.
}
              \label{fig.varyphi}
    \end{figure*}

We normalise distances by the characteristic length $L$,
the magnetic field by $B_0$, and times by the nonrelativistic
gyroperiod associated
with a magnetic field of magnitude $B_0$: $T=2\pi m_0 c /(|q|B_0)$.

The value of the electric field magnitude $E_0$ is an input parameter to
our numerical code. In order to compare the efficiency of acceleration
in fan and spine reconnection, we choose $E_0$ in such a way
that the average value of $|\vec{E}|$ in a random set of points
on a sphere of radius 1 centered
in the null point, is the same for the two cases.
We find that a value $E_0$=0.1025 statvolt/cm $\sim$3 kV/m corresponds to the
same average electric
field on the sphere as we used in our previous spine reconnection work [\cite{Dal2005,Dal2006}].
We choose the parameter $l$, defined in Section \ref{sec.config}, 
to have a value 10$^{-5}$: we verified that the same
results are obtained when a  value 10 times smaller is chosen.
We use $B_0$=100 gauss and L=10 km.
The magnetic field value used is typical of a solar Active Region.
The effect of varying these parameters will be described in detail elsewhere
[Dalla and Browning, in preparation, 2008]; the typical behaviour
is that decreasing the value of $B_0$ results in a faster drift of
particles towards the region of strong magnetic field and consequently 
a more efficient acceleration.

In Figure \ref{fig.varybeta} we show three particle trajectories 
with initial positions in the plane through the spine at $\phi$=100$^{\circ}$
and three values of the latitude $\beta$ (note that the fan plane corresponds
to $\beta$=0).
The distance from the
null, $r$=$(x^2+y^2+z^2)^{1/2}$, is $r$=1 at the initial time.
The initial particle energy is 300 eV.
There is a strong dependence of the energy gain on the initial value of 
the latitude, with particles starting near the fan plane gaining the most
energy. Particles starting at latitudes greater than about
$\beta$$\sim$30$^{\circ}$  
have very little acceleration.
While moving towards the fan plane, where the electric field is strong,
particles are at the same time moving away from the null point
and towards regions of stronger magnetic field, as shown in the $x$-$y$ projection.

In Figure \ref{fig.varyphi} we keep the latitude of the particle's 
initial position constant (equal to $\beta$=5$^{\circ}$), and vary the azimuthal angle.
Here we can see that an initial position at $\phi$=85$^{\circ}$ results in
stronger acceleration than one at $\phi$=65$^{\circ}$, and very little
acceleration
is achieved if the starting position is at 
$\phi$=45$^{\circ}$.
The trajectories of Figure \ref{fig.varyphi} can be explained by considering
the components of the electric drift velocity:
\begin{equation} \label{eq.ve}
\vec{v}_E= c \, \frac{{\vec{ E}} \times {\vec{ B}}}{B^2}
\end{equation} 
with $\vec{ E}$ and $\vec{ B}$ given by eqs.(\ref{eq.bfan1})--(\ref{eq.efan3}).
Calculating the $z$ component of $\vec{v}_E$ shows that it is largest
when  $\phi$=90$^{\circ}$ and it is zero at
 $\phi$=0$^{\circ}$. Hence azimuthal locations near $\phi$=90$^{\circ}$
result in more efficient drift toward the fan plane, where the electric
field is large and acceleration is most efficient. 

Figures \ref{fig.varybeta} and \ref{fig.varyphi} show that, unlike in
spine reconnection, there is strong drift in the azimuthal direction.
The only exception to this is the motion of a particle that starts
the $y$=0 plane (corresponding to $\phi$=90$^{\circ}$), for which the motion
remains planar. 
In this
plane only, the flow lines resemble those of the spine reconnection 
case, and particles can cross the fan plane and exit the region 
near the null at negative latitudes. In all other planes
through the spine, particles cannot cross the fan plane.

   \begin{figure*}
   \centering
   \includegraphics[width=11cm]{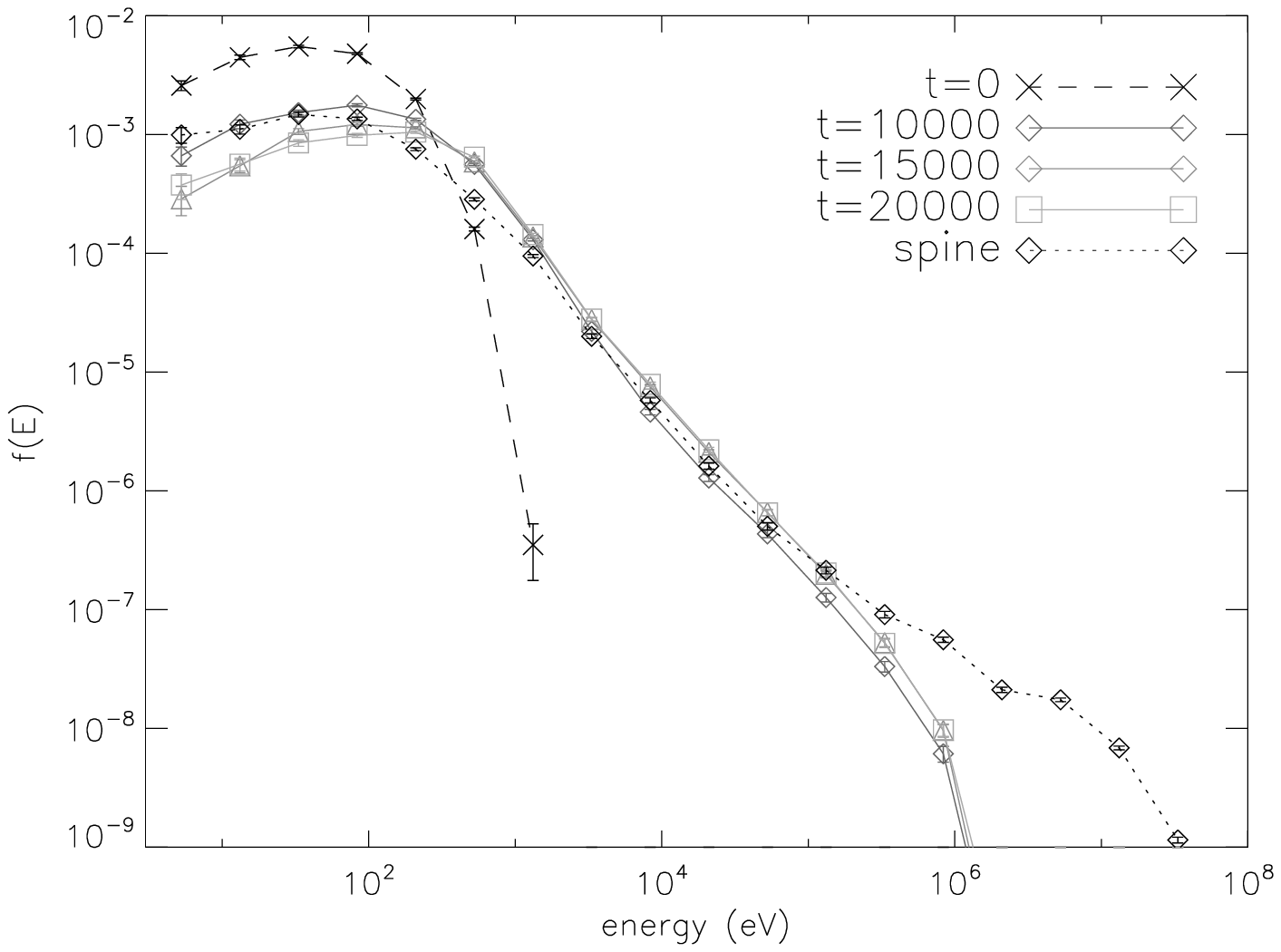}
   \caption{Time evolution of the energy spectrum during fan reconnection with
     $E_0$=3 kV/m, up to a final time $t$=20000. The dotted line gives the
     spine reconnection spectrum 
at $t$=10000. }
              \label{fig.fanspectrum}
    \end{figure*}

The $x$-$y$ projections in Figures  \ref{fig.varybeta} and \ref{fig.varyphi}\ show a tendency
for particles to move parallel to the $y$ axis once they have left  
the region near the null point. This can be understood by introducing
a potential $\psi$ such that ${\bf E}=-\nabla \psi$. From 
Eqs.~(\ref{eq.efan1})--(\ref{eq.efan3}) one obtains:
\begin{equation}
\psi(x,y,z)= E_0 \, L \, \frac{x \, \sqrt{|z|}}{(4+|z|\,y^2)^{1/2}} \label{eq.potential} 
\end{equation}
(here $\psi$ is dimensional while $x$, $y$ and $z$ are dimensionless).
Conservation of energy requires that the quantity $W$=$ K + q \psi $, 
where $K$ is the kinetic energy, is constant during the motion.
When particles are sufficiently far from the null point, their
kinetic energy remains approximately constant,  as can be seen in
the lower panels of Figure \ref{fig.varybeta} and Figure \ref{fig.varyphi}, therefore conservation
of energy implies that $\psi(x,y,z)$ must also remain 
constant.
For the trajectories
of Figure \ref{fig.varybeta}, for example, a constant kinetic energy is achieved for
$y$-values ranging from $-$0.2 (for $\beta$=60$^{\circ}$) to $-$3.9 (for
$\beta$=3$^{\circ}$) (note that the distance of the particle from the fan
plane is also a factor in determining when acceleration stops, because at
larger distances from this plane the electric field is less strong).
At the same time, the value
of $z$ is small and slowly decreasing: the values of $z$ at the time when the 
kinetic energy becomes constant for the Figure \ref{fig.varybeta} trajectories are $z$=0.0004, 0.0019, 0.010 and 0.18
for $\beta$=3$^{\circ}$, 10$^{\circ}$, 25$^{\circ}$ and 60$^{\circ}$ respectively.
From Eq.~(\ref{eq.potential}), we can see that a constant $\psi$
and very small $z$ give a tendency for
$x$ to remain roughly constant, or in other words
for the motion to be approximately parallel to the $y$ axis. This tendency is
stronger for particles that get nearer the fan plane: these are also those
that achieve largest acceleration. The direction of the $y$ axis is the preferential flow
direction within this type of fan reconnection, as described in Section \ref{sec.config}.

It is also interesting to look at the properties of the bulk flow velocity $V$
near the plane $z$=0, given, for $z$$>$0, by:
\begin{eqnarray}  
V_{x}  & \sim & \frac{x \, y}{4(x^2+y^2) \sqrt{z}} \label{eq.vx}\\
V_{y} & \sim & \frac{- x^2}{4(x^2+y^2) \sqrt{z}} \label{eq.vy}\\
V_{z} & \sim & \frac{- y \sqrt{z}}{2(x^2+y^2) }\label{eq.vz}
\end{eqnarray}  
Eqs.~(\ref{eq.vx})--(\ref{eq.vz}) are obtained from Eqs.~(4.25) of
  \cite{Pri1996}, by taking the limit $z$$\rightarrow$0 (with our equations having
opposite sign to \cite{Pri1996} to ensure that 
${\bf E}=-\nabla \psi$). 
Near the fan plane, the $V_{z}$ component
is small so that particles tend to remain near this plane.
The streamlines in the fan plane are approximately circles 
(see Figure \ref{fig.fanflow}-a). The drift velocity is generally directed
from the positive $y$ quadrant to negative $y$, with  a strong flow in the
negative $y$ direction across the $x$-axis; this also shows that  particles
tend to pick up a velocity anti-parallel to the $y$ axis.  Furthermore,
particles starting closer to the positive $y$ axis (at larger $\phi$) remain
in the swirling stream for longer time, and so gain more
energy (see Figure \ref{fig.varyphi}).

\section{Particle spatial distributions and spectra}\label{sec.manypart}

To analyse the evolution of a population of particles during fan reconnection,
we inject 10000 protons at random positions  
on a sphere of radius 1 around the null point, in the same 
way as done by \cite{Dal2006} for spine reconnection.

Figure \ref{fig.fanspectrum} shows the time evolution of the energy
spectrum, for 10000 protons with initial Maxwellian energy distribution
with temperature = 10$^6$K. We find that in order to reach a steady-state 
spectrum, we need to integrate particle trajectories up to a normalised time 
$t$=20000, approximately double the time that was required to
reach a steady-state during spine reconnection for the same parameters.
The dotted line shows the steady-state energy spectrum from the spine reconnection
analysis \cite{Dal2006}, at the final time $t$=10000.
Comparison of the fan and spine plots at the final times shows that 
spine reconnection produces a larger population of high energy particles
(protons above 0.1 MeV).

   \begin{figure*}
   \centering
   \includegraphics[width=14cm]{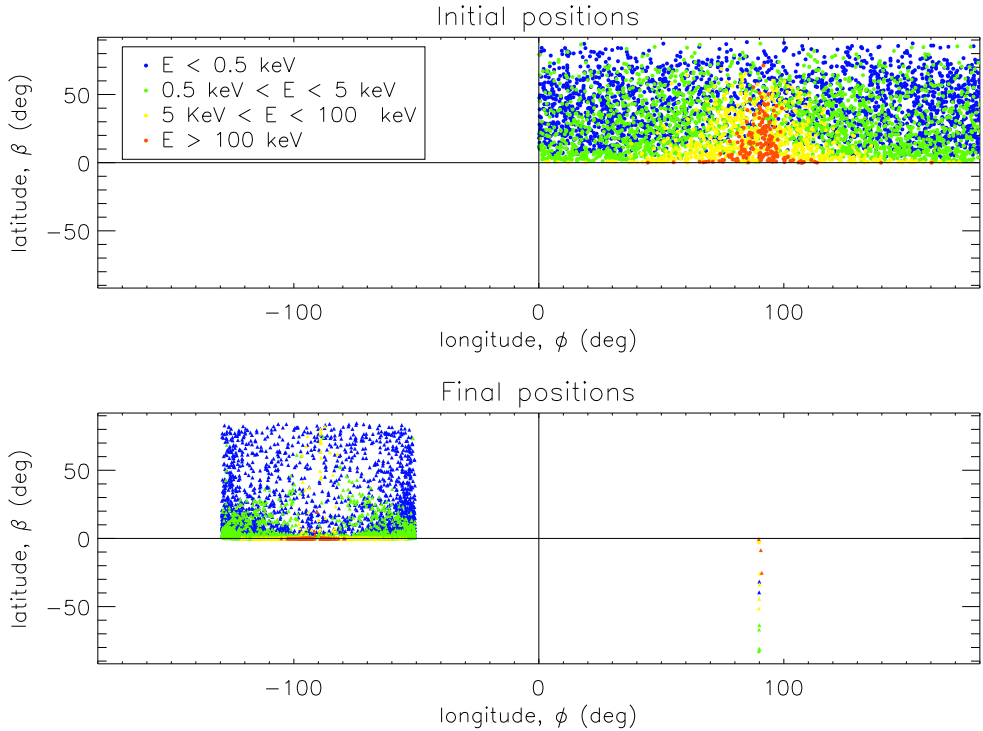}
   \caption{Initial and final positions of particles during fan reconnection
     at the initial time t=0 (top) and final time t=20000 (bottom).}
              \label{fig.fanpositions}
    \end{figure*}

   \begin{figure*}
   \centering
   \includegraphics[width=15.2cm]{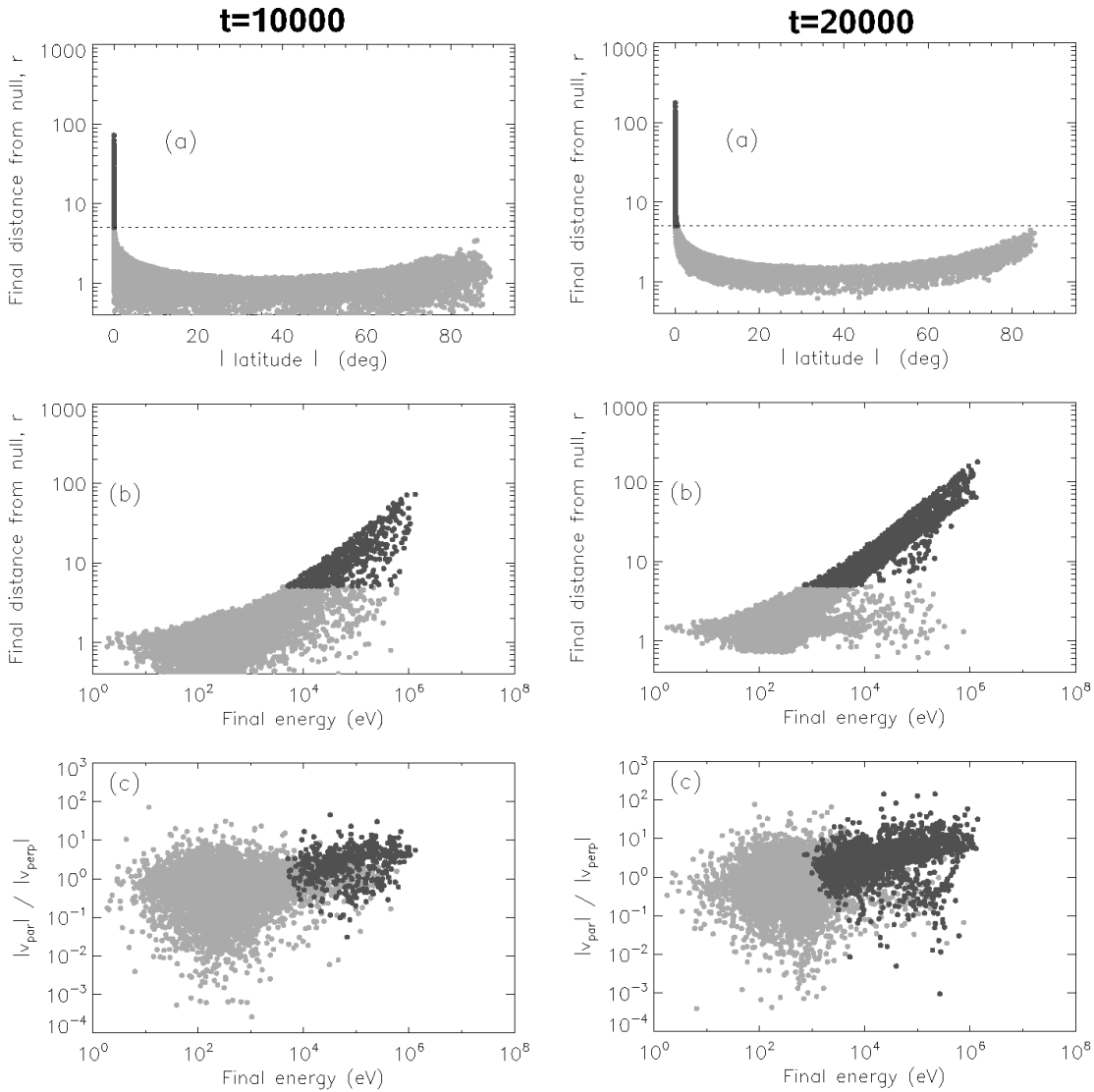}
    \caption{Parameters of the energetic particle populations during fan
    reconnection at $t$=10000 (left) and $t$=20000 (right). 
    {\it(a)} Distance from the null versus absolute value of latitude
    at the given time.
    {\it(b)} Distance from the null versus particle energy. 
    {\it(c)} Ratio of the magnitude of the component of velocity parallel to
    the magnetic field to perpendicular component, versus  particle energy.
    Dark grey points indicate particles with $r$$>$5 and small $|\beta|$, light grey points
    all the other particles.}
              \label{fig.fanescape}
    \end{figure*}

   \begin{figure*}
   \centering
   \includegraphics[width=7.9cm]{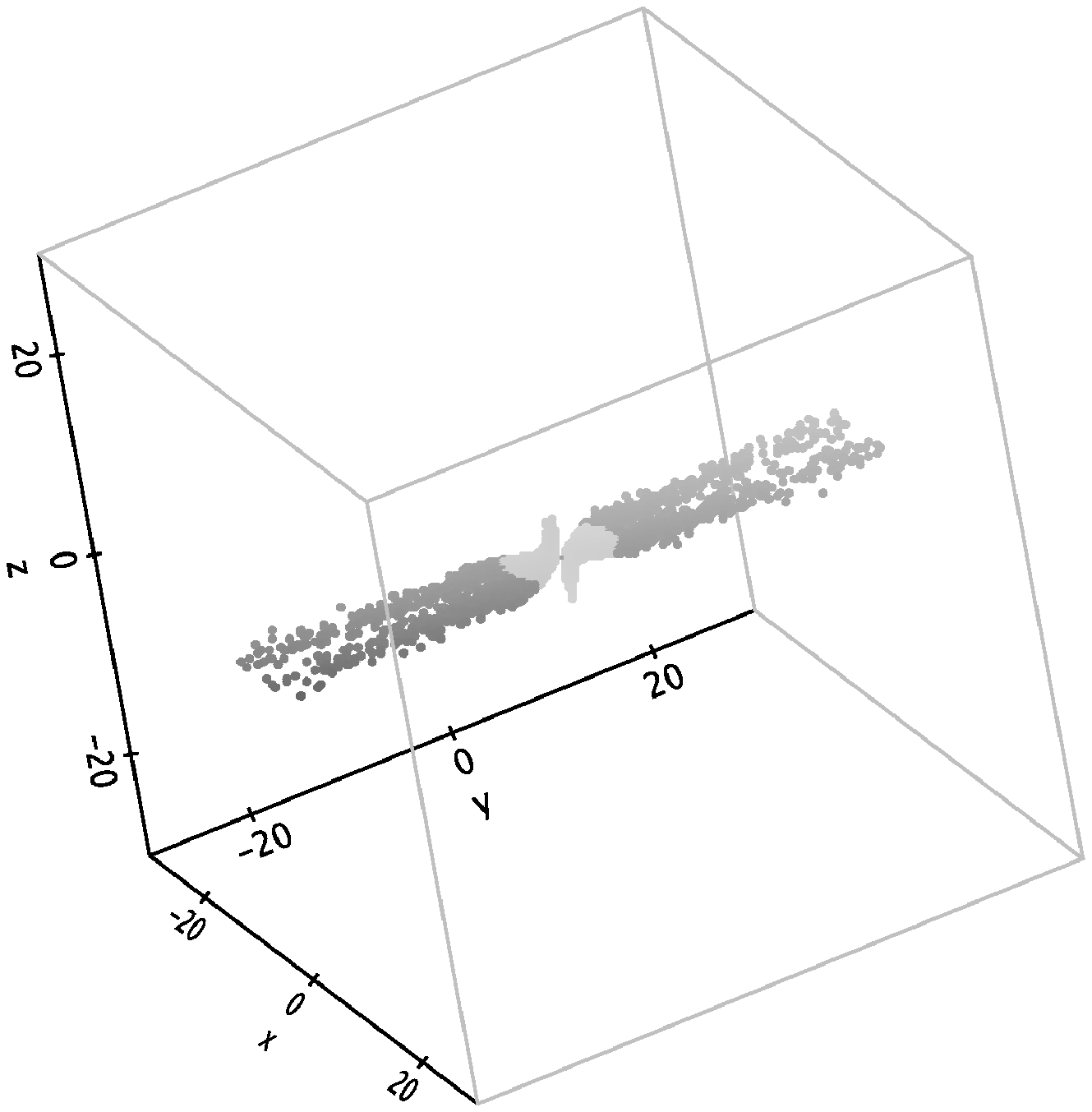}
    \caption{3D view of particles at t=20000, for particles within a cube of
      size 30 centered on the null. The colour coding is the same as in Figure
      \ref{fig.fanescape}, showing in dark grey particles with $r$$>$5. These 
      are are escaping in a ribbon-like 
     configuration. } \label{fig.ribbon}
    \end{figure*}

In the intermediate energy range  (protons 1 keV to 0.1 MeV), the efficiency
of fan and spine reconnection is similar once a steady state is reached.
The slope of the curve is very similar in this energy range, hence so is the 
power law index of the spectra.

Figure \ref{fig.fanpositions} shows the angular location of the particles in our population 
at the initial time (top plot) and the final time (bottom
plot). Each location is identified by its latitude $\beta$, i.e. 
the elevation angle from the $x$-$y$ plane, and longitude $\phi$.
Inflow regions are the top-right and bottom-left quadrants in the
$\phi$, $\beta$ representation.
In Figure \ref{fig.fanpositions} we only show the initial and final positions of
particles starting in the top-right inflow region. 
Particles (not shown) starting in the bottom left quadrant have
distributions of initial and final positions 
symmetric with respect to those shown in Figure \ref{fig.fanpositions}, 
apart from statistical
fluctuations. Each point of Figure \ref{fig.fanpositions} represents one
particle and is colour coded according to the final 
particle energy. It should be noted that the energy ranges corresponding to
the various colours are different from
those used in \cite{Dal2006}, since the maximum particle energies obtained
in fan reconnection are lower than in spine reconnection.

Figure \ref{fig.fanpositions} shows that particles starting near the
fan plane are those that gain the largest energy, as would be expected since
the fan plane is the location where the electric field is largest.
As one moves towards increasing latitudes, particles will reach the
fan plane with greater difficulty and gain less energy.
Figure  \ref{fig.fanpositions} shows that at the final time, particles are
confined to a range of longitudes [$-$130$^{\circ}$,$-$50$^{\circ}$] (and symmetrically for negative
latitudes, not shown in Figure  \ref{fig.fanpositions}, longitudes [50$^{\circ}$,130$^{\circ}$]).
This is a result of the tendency of particles to move anti-parallel to
  the $y$ axis after they
have passed the region near the null, as discussed in
Sec.~\ref{sec.singlepart}. This is clearly shown in
Figure \ref{fig.varyphi}, in which particles with initial positions covering 
a wide range of azimuthal angles $\phi$, leave the null moving anti-parallel to
the $y$ axis and within a narrower band of $\phi$. 

Figure  \ref{fig.fanpositions} also shows that initial azimuthal locations 
near $\phi$=90$^{\circ}$ are the most favourable for particle acceleration since
near this plane the $z$ component of the electric field drift, pushing
particles towards the fan plane, is largest. Furthermore, such
particles spend longest in the swirling flows, as shown in Figure \ref{fig.varyphi}.

In \cite{Dal2006} we reported that the spine reconnection configuration
naturally produces two symmetric jets of energetic particles escaping 
along the spine. The escaping particles in the two jets were found at
large distances from the null point at the final time.

Figure \ref{fig.fanescape} adds information about the distance from the 
null point at the final time, to the angular information displayed
in Figure  \ref{fig.fanpositions}.
Figure \ref{fig.fanescape} displays a snapshot of the same information 
at $t$=10000 (left panels) and  $t$=20000 (right panels). Panels (a)
show the distance from the null point versus magnitude of the latitude $\beta$. 
No particles are found at
values of $|\beta|$ near 90$^{\circ}$, i.e. no particles are escaping
along the spine during fan reconnection.

However some particles are escaping from the configuration after acceleration:
they are represented in Figure \ref{fig.fanescape} by the dark grey points
found at low $|\beta|$, i.e. near the fan plane. The same particles found
around $r$=50 at $t$=10000 are found at approximately   $r$=100 at $t$=20000.
These particles have energies in the range 10 keV to 1 Mev at  $t$=20000
(panel (b))
and are characterised by a ratio of the parallel to perpendicular component of
velocity greater than 1 (panel (c)). Consequently they are escaping from the
magnetic configuration.
 
Figure \ref{fig.ribbon} gives a 3D view of the escaping population, showing
that they are moving away from the magnetic null in a ribbon like structure.
The orientation of the ribbon is parallel to the direction along which
the fan reconnection flows have been imposed (the $y$ axis in our reference
system).
%

\section{Conclusions}

In this work we analysed test particle trajectories in the fields
characteristic of fan reconnection, one of the two regimes of
magnetic reconnection at a 3D magnetic null.
We aimed to investigate how efficiently particles can reach the
region of strong electric field where acceleration is possible,
and the geometry of particle escape after acceleration has
taken place.
Also, we assessed the differences between the fan regime and
the previously studied spine reconnection one.

We found that the fan reconnection regime is less efficient in
seeding test particles to the region of strong electric field,
compared to the spine reconnection one. The time
required for acceleration to take place is approximately twice
the one we previously found for the spine regime [\cite{Dal2006}].
This is due to the flow patterns for fan reconnection making
it difficult for particles starting at high latitude locations
to reach the fan plane, where the electric field is large.

The steady-state spectrum we obtained for fan reconnection has 
a spectral index in the intermediate energy range (protons 1 keV to
0.1 MeV, for the parameters used in our study)
that is similar to the spine case one. However, within our model, it appears 
that fan reconnection is less efficient in the generation of
a very high energy population (protons $>$0.1 MeV), compared to 
the spine case.
It should be emphasised that our model, based on the
\cite{Pri1996} formulation, does not accurately represent the small region
where resistive (or other dissipative) effects become important. Our statement regarding the
maximum particle energy produced by each of the two regimes will
need to be re-assessed within a model including parallel electric fields within 
the resistive region, and which properly models the "cut off" of
the ideal electric field as this region is approached. This will be the subject of future study.
However, our model does properly represent the inflow
of particles to the dissipative region where parallel electric fields
exist. Hence,
only those particles reaching the fan plane (the high energy tail in our
calculation) may be further accelerated to even higher energies. We thus
expect that using  a more self-consistent reconnection model
will affect only the high energy tail
and cut-off of the energy spectra.

Regarding the geometry of escape from the region near the null after
acceleration, we found that in the fan reconnection case energetic
particles escape along the fan plane in a ribbon-like configuration.   
We do not observe particle escape along the spine line, as was the
case in the spine regime [\cite{Dal2006}]. Thus, in principle,
observations of the spatial distribution of high energy particles 
might be a discriminator between spine and fan reconnection.

\begin{acknowledgements}
We thank the referee for useful comments that have improved the paper.
We acknowledge financial support from PPARC.
\end{acknowledgements}

\end{document}